# High pressure effect on superconductivity of YB$_6$
Revised 7/17/2014  08:49:00

S. Gabáni,[1,*] I. Takáčová,[1] G. Pristáš,[1] E. Gažo,[1] K. Flachbart,[1] T. Mori,[2] D. Braithwaite,[3] M. Míšek,[4] K.V. Kamenev,[4] M. Hanfland,[5] and P. Samuely[1]

[1]*Centre of Low Temperature Physics @ Institute of Experimental Physics, Slovak Academy of Sciences, Watsonova 47, 040 01 Košice, Slovakia*
e-mail address: gabani@saske.sk
[2]*Advanced Materials Laboratory, National Institute for Materials Science, Namiki 1-1, 305 0044 Tsukuba, Japan*
[3]*Université Grenoble Alpes, INAC- SPSMS, 17 rue des Martyrs, 38000 Grenoble, France*
*CEA, INAC-SPSMS, 17 rue des Martyrs, 38000 Grenoble, France*
[4]*Centre for Science at Extreme Conditions, University of Edinburgh, Mayfield road, EH9 3JZ Edinburgh, United Kingdom*
[5]*European Synchrotron Radiation Facility, 6 Rue Jules Horowitz, 38000 Grenoble, France*

Pressure effect on superconducting properties of two YB$_6$ samples ($T_c$ = 5.9 and 7.5 K) were investigated by measurements of electrical resistivity, magnetic susceptibility, and X-ray diffraction in the pressure range up to 320 kbar. Magnetoresistivity measurements down to 60 mK and up to 47 kbar have shown a negative pressure effect on $T_c$ as well as on the third critical field $H_{c3}$ with the slopes $d\ln T_c/dp$ = -0.59%/kbar and $d\ln H_{c3}/dp$ = -1.1%/kbar, respectively. The magnetic susceptibility measurements evidenced that the slope of $d\ln T_c/dp$ gradually decreases with pressure reaching 3 times smaller value at 112 kbar. The lattice parameter measurements revealed the volume reduction of 14% at 320 kbar. The pressure-volume dependence is described by the Rose-Vinet equation of state. The obtained relative volume dependence $d\ln T_c/d\ln V$ analyzed by the McMillan formula for $T_c$ indicates that the reduction of the superconducting transition temperature is mainly due to hardening of the Einstein-like phonon mode responsible for the superconducting coupling. This is confirmed by the analysis of the resistivity measurements in the normal state up to $T$ = 300 K performed at pressures up to 28 kbar.

**Keywords:** YB$_6$, superconducting phase diagrams, electron-phonon interaction, high pressure effects

**PACS:** 74.70.Ad, 74.25.Dw, 74.62.Fj, 74.25.F-, 41.60.Ap

## I. INTRODUCTION

Among the large number of boron-rich binary compounds MB$_x$ (x ≥ 6) [1, 2], superconductivity has been found in only eight systems: MB$_6$ (M = Y, La, Th, Nd) and MB$_{12}$ (M = Sc, Y, Zr, Lu) [3, 4]. Among these, yttrium hexaboride exhibits the highest transition temperature $T_c$ reaching 8 K [5]. YB$_6$ crystallizes in the *bcc* CaB$_6$-type structure (space group *Pm3m*), in which the B$_6$ molecule has the octahedral form. In recent years, properties of YB$_6$ have been intensively investigated by specific heat, resistivity, magnetic susceptibility, and thermal expansion measurements [5, 6], optical [7], NMR [8], μSR experiments [9], point contact spectroscopy [10], as well as by means of electronic structure calculations [11, 12]. An important issue was to explain significant differences in $T_c$ between materials with otherwise very similar electronic and lattice properties. A conclusion could be deduced that the superconductivity of YB$_6$ is mediated mainly by the very soft phonon mode located at ≈ 7.5 meV [5, 10] and originating from the rattling motion of the Y ion in the spacious cage of the B$_6$ octahedron while the boron phonons are less important.

Lattice compression can affect all important constituents of superconductivity, namely phonon frequencies, the electron-phonon coupling constant $\lambda$, as well as the electronic density of states EDOS. To our knowledge there has been only a single experimental study of the pressure effect on the superconducting properties of YB$_6$ [13]. In particular, the negative effect on the transition temperature $T_c$ has been found for pressures up to 9.2 kbar with $dT_c/dp$ = -0.055 K/kbar. The upper critical field was suppressed as $dH_{c2}/dp$ = -4.84 mT/kbar showing an increase of the coherence length. Since no pressure effect on the penetration depth was observed, the Ginzburg-Landau parameter $\kappa$, which is the ratio of the penetration depth and the coherence length, was found decreasing and driving YB$_6$ towards the type-I superconductivity with increasing pressure.

Xu *et al.* [12] performed extensive *ab initio* studies of the effect of pressure on the electronic, vibrational, and superconducting properties of YB$_6$ in a wide range of pressures up to 400 kbar. Their calculations of the electron-phonon interaction $\alpha^2F(\omega)$ show a dominancy of a low-lying Einstein-like vibrations of Y atoms at about 8 meV, which makes about 86% of the electron-phonon coupling. The pressure effect on $T_c$ is negative. At $p$ = 400 kbar, the Y low-lying phonon mode still determines the superconducting coupling, but due to its hardening the coupling constant $\lambda$ and subsequently $T_c$ get significantly smaller. No experimental data in the very high pressure range have been available. Our study brings results of such an experiment.

The paper is organized as follows. After Introduction and Experiment the section III. Results and discussion follows. Our results on the influence of a hydrostatic pressure on magnetoresistivity up to 47 kbar are presented in Section III.1. showing a negative pressure effect on $T_c$ and $H_{c3}$. In Section III.2 the measurements of magnetic susceptibility up to 112 kbar and lattice parameter up to 320 kbar present how the $T_c$-reduction rate is slowed down at higher pressures. The *pressure versus volume* dependence allows to estimate the volume changes of $T_c$ and make an analysis. Precise zero field resistivity measurements in the normal state between 10 and 300 K and up to 28 kbar in Section III.3 were used to reveal a significant mode of the electron-phonon interaction and its variations with increasing pressures, which lead to suppression of superconductivity.

## II. EXPERIMENT

Two single crystals with different transition temperatures (sample #1 with $T_c$ = 7.5 K and sample #2 with $T_c$ = 5.9 K) of YB$_6$ were grown by the rf-heat floating zone method in an atmosphere of argon at a pressure of 5 bar. The residual resistivity ratio of the sample #1 was $RRR$=4.3 with the low temperature resistivity $\rho(0)$=10 $\mu\Omega$cm while for the sample #2 $RRR$=2.5 and $\rho(0)$=21 $\mu\Omega$cm. As shown by Lortz *et al.* [5], the $T_c$ value is related to the Y/B ratio, the higher $T_c$ corresponding to lower boron concentration with $T_c$ = 7.6 K for YB$_{5.7}$ and $T_c$ = 6.6 K for YB$_{5.9}$. The high pressure magnetotransport experiments were performed in a piston cylinder cell - PCC ($p \leq 28$ kbar) and in a diamond anvil cell - DAC ($p \geq 30$ kbar) (IEP Košice). Daphne oil or liquid argon as pressure transmitters and Pb or ruby fluorescence manometers were used in the PCC and DAC, respectively. The actual pressure upon loading was determined at room temperature, the pressure change on cooling is estimated to be less than 2 kbar. The temperature and magnetic field dependences of the resistivity between 2 and 300 K were measured by means of a PPMS instrument (Quantum Design) and in a home-built dilution $^3$He-$^4$He minirefrigerator below 2 K down to 60 mK.

The *ac*-magnetic susceptibility was measured in a DAC using the technique described in [14], where a pick-up coil consisting of about 10 turns of 12 $\mu$m diameter copper wire, which is inserted in the sample chamber (CEA Grenoble). The primary coil placed outside the sample chamber produced an excitation field of about 1 Oe at a frequency of 653 Hz. In this case the pressure was varied and determined in-situ at low temperatures with an accuracy better than 1 kbar.

The *dc*-magnetic susceptibility measurements were performed in CSEC Edinburgh using a miniature high-pressure cell for a MPMS instrument (Quantum Design) with a superconducting quantum interference device (SQUID) magnetometer using a *dc* field of 100 Oe [15]. Daphne 7373 oil was used as the pressure-transmitting medium.

Pressure dependence of the lattice parameter of YB$_6$ at room temperature was obtained by X-ray diffraction experiments up to 320 kbar at the ID09A beamline in the European Synchrotron Radiation Facility in Grenoble. Boheler Almax-type of DAC with a Re gasket and He as the pressure transmitter were used for the diffraction experiments [16].

## III. RESULTS AND DISCUSSION

### 1. Magnetotransport experiments at pressure

On the sample #1 the magnetization and specific heat measurements in fields up to 1 Tesla at ambient pressure were performed previously, from which the lower critical field $H_{c1}$ = 36 mT, the thermodynamical critical field $H_c$ = 72 mT and the upper critical field $H_{c2}$ = 280 mT have been determined [6]. They are in a very good agreement with the data of Lortz et al. [5] obtained on YB$_6$ sample with very similar $T_c$ = 7.2 K and $RRR$=3.9. The Ginzburg-Landau coherence length $\xi$ = 34 nm is obtained from our upper critical field $H_{c2}(0)$. From the residual resistivity $\rho(0)$ = 10 $\mu\Omega$ cm within the free electron model the electronic mean free path of about 20 nanometers follows. If the Pippard coherence length is calculated from the ratio of the Fermi velocity $v_F \approx 10^5$ m/s [9] and the superconducting energy gap $\Delta$ = 1.2 meV [10], one obtains $\xi_0 = (\hbar v_F)/(\pi\Delta) \approx 20$ nm of the same order as the mean free path indicating that the samples are close to the transition between the clean and dirty limit.

Now the resistivity measurements in magnetic fields at different pressures were performed on the same sample #1 in a configuration suitable to measure the third critical field, $H_{c3}$, i.e. with the current and voltage probes placed on the surface of the sample parallel with the applied magnetic field. The critical field $H_{c3}$ or the transition temperature $T_c(H)$ has been determined from the magnetoresistive superconducting transitions at its steepest slope, which is around 50 % of the normal state resistance during the temperature or field sweeps. Representative resistivity measurements on sample #1 has recently been presented in [17]. The obtained ambient pressure zero-temperature value of $H_{c3}(0)$ = 450 mT is to be compared with $H_{c2}(0)$ giving the ratio $H_{c3}(0)/H_{c2}(0)$ = 1.6 which is quite close to the theoretical prediction of 1.695. Again, our resistively determined third critical field at ambient pressure is very close to the one determined by Lortz et al. As the third critical field is directly proportional to $H_{c2}$, in the following we assume the same pressure and temperature dependence of the both quantities $H_{c3}$ and $H_{c2}$.

Figure 1 shows by different symbols, the resulting temperature dependences of $H_{c3}(T)$ at pressures of 4.5, 14, 22, 30, and 47 kbar generated in PCC and DAC. The zero-pressure and 9 kbar curves are not shown as they largely overlap with the curves at closest pressures. The graph reveals a systematic decrease of the zero-field transition temperature $T_c$ as well as of the zero-temperature value of $H_{c3}$ with increasing pressure.

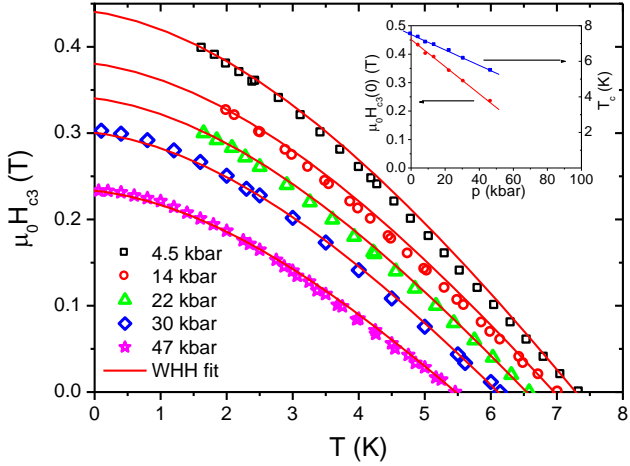

FIG. 1. Temperature dependences of the third critical field, $H_{c3}$, at different pressures for sample #1. Lines represent WHH fits. Inset: linear decrease of $H_{c3}$ and $T_c$ with increasing pressure.

As documented in Fig. 1, the $H_{c3}(0)$ values could be extrapolated using the Werthamer-Helfand-Hohenberg (WHH) fits [18]. The $H_{c3}(T)$ curves taken at different pressures are not parallel, but their slope near $T_c$, $(dH_{c3}/dT)_{T=Tc}$, systematically decreases with increasing pressure by $d(dH_{c3}/dT)_{T=Tc}/dp = -0.6$ mT K$^{-1}$/kbar. This is related to the fact that pressure effect on the upper critical field is significantly stronger than that on $T_c$. Both quantities decrease linearly in the measured pressure range (see the inset of Fig. 1). The linear fits give $dT_c/dp = -0.044$ K/kbar. Note that this value is smaller than $dT_c/dp = -0.055$ K/kbar obtained by Khasanov et al. [13] on a sample with $T_c = 6.6$ K, but the latter was measured only up to 9.2 kbar and as we will show below the effect of pressure on $T_c$ weakens at higher pressures. If we make a linear fit only to the data points at pressures below 10 kbar the resulting $dT_c/dp$ slope is even closer to the value of Khasanov. The relative change of the transition temperature with pressure is $d\ln T_c/dp = -0.59$ %/kbar. The linear fit of the pressure dependence of $H_{c3}$ yields $dH_{c3}/dp = -4.64$ mT/kbar and the relative change $d\ln H_{c3}/dp = -1.1$%/kbar. The pressure, at which the critical temperature should be suppressed to zero, can be estimated $p_c \approx 170$ kbar using the linear extrapolation.

The almost twice larger relative decrease of the critical field with pressure ($d\ln H_{c3}/dp = -1.1$%/kbar) compared to the relative decrease of the critical temperature with $d\ln T_c/dp = -0.59$ %/kbar and the relative decrease of the temperature derivation of the third critical field at $T_c(0)$ with $d\ln\{(dH_{c3}/dT)_{T=Tc}\}/dp = -0.65$ %/kbar can be explained as follows. The upper critical field is given by the relation $H_{c2} = \Phi_0/(2\pi\xi^2)$, where $\Phi_0$ is the magnetic flux quantum and $\xi$ the coherence length. The relative pressure change of the upper critical field can be written as $d\ln H_{c2}/dp \propto -2d\ln\xi/dp$. The coherence length obeys the relation $\xi(0) = (\hbar v_F)/(\pi\Delta)$ and $2\Delta \propto k_B T_c$ so then, $d\ln\xi/dp = d\ln v_F/dp - d\ln T_c/dp$. If the pressure change of the Fermi velocity $v_F$ is negligible in comparison with the change of $T_c$, we will get the relation $d\ln H_{c2}/dp \propto 2d\ln T_c/dp$ in agreement with our experimental findings. In the free-electron model the Fermi velocity $v_F \propto V^{1/3}$, where $V$ is the sample volume. The pressure change of the Fermi velocity is related to the bulk modulus $B_0$ as $d\ln v_F/dp \propto -1/3 d\ln V/dp = 1/3 B_0^{-1}$. The bulk modulus for YB$_6$ is $B_0 \approx 1700$ kbar (see Section III.2) giving a relative change of Fermi velocity $d\ln v_F/dp = 0.02$ %/kbar, which is indeed much smaller than the relative change of $T_c$. As has been already stated by Khasanov et al. [13], in superconductors for which $d\ln T_c/dp \gg 1/B_0$, the relative pressure change of the superconducting quantities such as $T_c$, $\xi(0)$, $H_{c2}$, but also $\kappa$ and $(dH_{c2}/dT)_{T=Tc}$ are not independent but related to each other like shown above. Our measurements have proven the validity of this relation for YB$_6$ in the pressure range up to almost 50 kbar.

Xu et al. [12], who performed extensive ab initio studies of the effect of pressure on the electronic, vibrational and superconducting properties of YB$_6$ up to 400 kbar, predict a negative pressure effect on $T_c$ with a coefficient of $dT_c/dp = -(0.024 \div 0.027)$ K/kbar in the pressures up to 200 kbar, what is approximately twice smaller than observed by us. Above 200 kbar the same theory predicts much slower decrease of $T_c$ with steepness of $dT_c/dp = -(0.003 \div 0.011)$ K/kbar. As shown above, the slope $dT_c/dp$ observed up to 47 kbar in our study is smaller than that of Khasanov et al. [13] found over a limited pressure range. All this indicates that the pressure effect on $T_c$ weakens at higher pressures. To check this quantitatively, the measurements of $T_c$ over a broader pressure range were carried out.

### 2. Magnetic susceptibility experiments and lattice parameter under pressure

Temperature dependencies of the magnetic susceptibility, $\chi(T)$, were carried out on the sample #1 and #2 by the dc-method ($\chi_{dc}$) to 100 kbar and by the ac-method ($\chi_{ac}$) up to 112 kbar, respectively. The critical temperatures at various pressures were associated with the steepest slope of $\chi(T)$ appearing, which is around 50 % of the $\chi(2 K)$ value.

Figure 2 depicts the *ac*-measurements. As seen from Fig. 2b, the resulting pressure dependences of the critical temperature $T_c$ show clearly a non-linear behavior for both samples in the measured pressure range. The *ac*-measurements exhibited less scatter, which allowed to determine for both samples the initial slope of $dT_c/dp = -0.044$ K/kbar in the range below $\approx 50$ kbar. In the highest pressure range, above about 90 kbar, the slope changes to $dT_c/dp \approx -0.015$ K/kbar. Our measurements bring a natural explanation of the higher $dT_c/dp$ values obtained by Khasanov *et al*. in the limited pressure range (see our first three points of the sample #2) and also support the calculations of Xu *et al*.

of the $V(p)$ dependence. It can be well fitted by the Rose-Vinet equation of state [19]:

$$p = 3B_0 \left( \frac{1-v}{v^2} \right) \exp\left[ \frac{3}{2}(B_0' - 1)(1-v) \right], \quad (1)$$

where $v = (V/V_0)^{1/3}$, $B_0$ is the bulk modulus and $B_0'$ its derivative. The obtained value of the bulk modulus, $B_0 = 1659$ kbar, is quite close to the value $B_0 \approx 1790$-$1900$ kbar, estimated from experiments in Ref. 20 or to those deduced from band-structure calculations of YB$_6$ [12, 21].

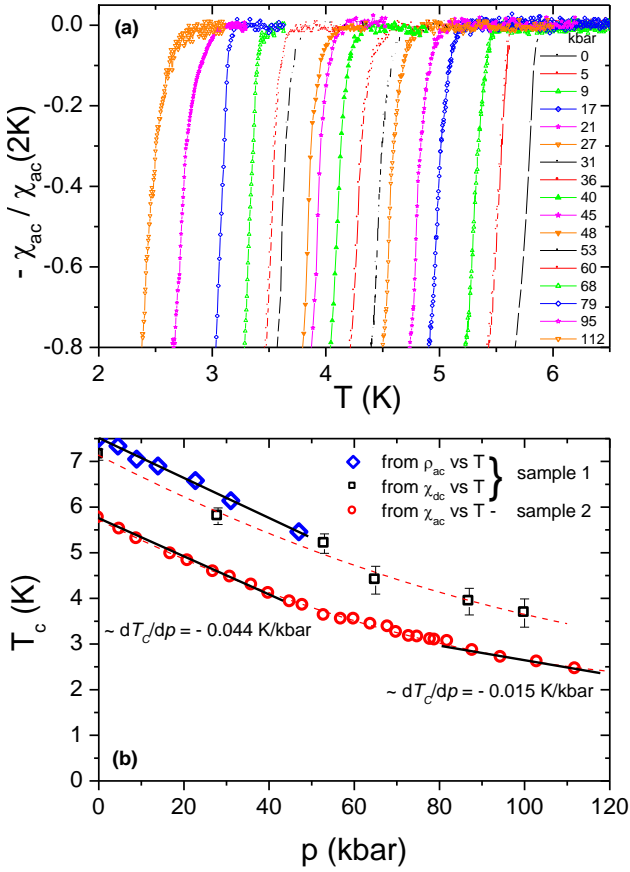
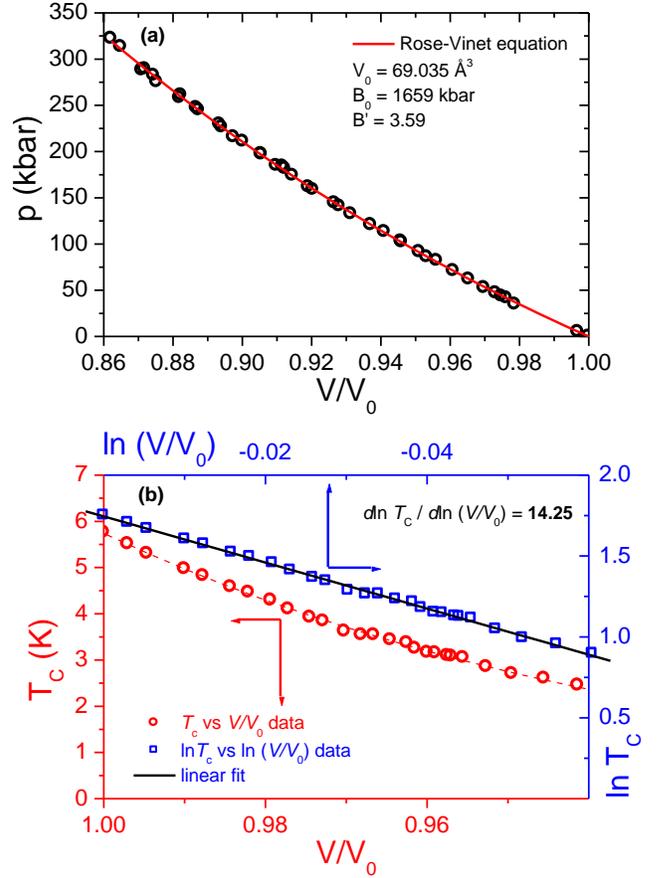

FIG. 2. Temperature dependences of the *ac*-susceptibility up to 112 kbar for sample #2 (a); and phase diagram, $T_c$ vs $p$, for sample #1 (received from $\rho_{ac}(T)$ and $\chi_{dc}(T)$ measurements) and sample #2 (received from $\chi_{ac}(T)$ measurements) (b). Lines represent linear fits. Dashed lines are guides for the eye.

The results of the X-ray diffraction study of the lattice parameter/volume and their development with pressure of the sample #2 are summarized in Fig. 3. The unit-cell volume is reduced by $\Delta V/V_0 \approx 14$ % at 320 kbar with $V_0$ being the volume of the unit cell at ambient pressure. Over the extended pressure range one can observe a non-linearity

FIG. 3. Pressure-volume relation of the YB$_6$–sample #2 at room temperature and up to 320 kbar (a), where line represents fit by Rose-Vinet equation of state (see text). Relative volume dependence of $T_c$ for the YB$_6$–sample #2 (circles), i.e., $T_c$ vs $V/V_0$ (left-bottom axis) and also $\ln T_c$ vs $\ln(V/V_0)$ (right-up axis) representation (b). Solid line represents the linear fit with slope of $d\ln T_c/d\ln(V/V_0) = +14.25$. Dashed line is a guide for the eye.

With the $p$-$V/V_0$ relation we can construct the volume dependence of the transition temperature $T_c$ taking the data from Fig. 2 for the sample #2. As shown in Fig. 3b the $T_c$ dependence on volume still remains non-linear (lower curve). The relative volume dependence of the critical temperature $T_c$ can be read from the upper curve giving an unusually large number of $d\ln T_c/d\ln(V) = 14.25$. Such a number was predicted from the analysis of the temperature dependence of the linear expansion coefficient [5] but it is first time directly experimentally determined here.

With those information we can try to analyze the origin of the negative pressure effect on the transition temperature in YB$_6$. In the most of superconductors, the pressure suppresses $T_c$ due to the stiffening of the lattice which weakens the electron-phonon interaction, but the changes in the electronic density of states can be in the game and in some cases $T_c$ can even increase [22-24].

The superconducting critical temperature can be estimated from the well-known Allen-Dynes modified McMillan formula comprising the most important superconducting parameters [25]:

$$T_c = \frac{\omega_{\ln}}{1.2} \exp\left[-\frac{1.04(1+\lambda)}{\lambda - \mu^*(1+0.62\lambda)}\right] \qquad (2)$$

The formula is valid for strong coupling superconductors ($\lambda \geq 1$) and connects the value of $T_c$ with the electron-phonon coupling constant $\lambda$, the logarithmically averaged phonon frequency $\omega_{ln}$, and the screened Coulomb repulsion parameter $\mu^*$.

Taking the logarithmic volume derivative of both sides of Eq. (2) and neglecting the volume dependence of $\mu^*$, which has little sensitivity to applied pressure, we obtain the simple relation [5, 26]:

$$\frac{d\ln T_c}{d\ln V} = -\gamma_{ph} + f(\lambda, \mu^*)\frac{\partial \ln \lambda}{\partial \ln V}, \qquad (3)$$

where $\gamma_{ph} \equiv -\partial\ln\omega/\partial\ln V$ is the Grüneisen parameter representing the anharmonicity of the lattice vibrations with the circular frequency $\omega$ and $f(\lambda, \mu^*) = 1.04\lambda[1+0.38\mu^*]/[\lambda-\mu^*(1+0.62\lambda)]^2$. In the case of YB$_6$ with $\lambda \cong 1.04$, $\mu^* \cong 0.1$ [5], $f(\lambda, \mu^*)$ is about 1.5. Lortz et al. [5] obtained the Grüneisen parameter from their thermal expansion experiment as $\gamma_{ph} \approx 9$. Then, using this value with our $d\ln T_c/d\ln(V) = 14.25$, we obtain $\partial\ln\lambda/\partial\ln V \cong 15.5$, which is again a very large volume dependence of the electron-phonon coupling constant.

If we express the electron-phonon coupling constant as $\lambda = \eta/M\omega^2$, where $\eta = N_E\langle I^2\rangle$ is the Hopfield electronic parameter (comprising the electronic density of states at the Fermi energy $N_{EF}$ and the mean square electron-ion matrix element $\langle I^2\rangle$) and the ionic mass $M$ then, it follows:

$$\frac{\partial \ln \lambda}{\partial \ln V} = \frac{\partial \ln \eta}{\partial \ln V} + 2\gamma_{ph} \qquad (4)$$

and we see that $\partial\ln\eta/\partial\ln V$ gives only a small contribution ($\cong -2.5$) to the overall change of $\lambda$, which is mostly determined by the Grüneisen parameter. Similarly, the relative volume change of the superconducting temperature is approximately $d\ln T_c/d\ln(V) \approx 2\gamma_{ph}$, meaning that the main reason for decrease of $T_c$ is indeed the hardening/unharmonicity of the relevant phonon mode. In the following section we address this issue experimentally.

## 3. Pressure effect on low energy mode responsible for superconducting coupling

Lortz et al. [5] exploited the normal state resistivity measurements as a "thermal" spectroscopy to deconvolve the spectrum of the electron-phonon interaction $\alpha^2F(\omega)$ and the coupling constant $\lambda$ in YB$_6$. Such an approach was successful because YB$_6$ happened to be an example of the superconductor with a dominant low energy Einstein-like phonon mode, which is well expressed in the temperature dependence of the normal state resistivity, heat capacity and thermal expansion as a hump or peak at about 50 K. Obviously, the "thermal" spectroscopy could not resolve the whole spectrum with modes stretching to energies up to almost 200 meV, but was capable to determine the most important low-energy part. The deconvolved spectrum $\alpha^2F(\omega)$ have recently been verified by our point-contact spectroscopic measurements at ambient pressure [10] and we found indeed that the dominant mode in the electron-phonon interaction mediating superconductivity in YB$_6$ is located at 7.5 meV. By employing point-contact spectroscopy under pressure [27] it should be possible to observe the shift of Y-phonon mode directly.

Here, we have used the same approach as Lortz et al. to analyze the temperature dependences of the normal state resistivity of the sample #2 measured at pressures 1 bar, 8 kbar, 17 kbar, 22 kbar, and 28 kbar. Figure 4a presents the raw data measured between 10 K and 300 K. As can be seen, the residual resistivity $\rho(0) = 21$ μΩ cm is almost invariable with pressure, but the high-temperature resistivity is reduced. In Fig. 4b, where the temperature derivative of the resistivity at ambient pressure is shown, makes the effect of low-energy modes of the electron-phonon interaction more visible showing a clear maximum. Figure 4c displays the changes of $d\rho/dT$ with pressure. Unfiltered raw data lead to rather noisy derivative plots. That is why only the $d\rho/dT$ data taken at 1 bar and 28 kbar are presented (symbols), while the fits by solid lines are shown for all pressures. The main result is well documented, namely that the maximum of $d\rho/dT$ is clearly shifting to higher temperatures by several kelvins (see the insert of Fig. 4c) at $p = 28$ kbar.

The temperature dependence of the resistivity in YB$_6$ is described by the Bloch-Grüneisen (BG) theory of the electron-phonon interaction [28]. Generally, the BG theory predicts a $\rho \sim T^5$ dependence at low temperatures and the linear temperature dependence at high temperatures, but in YB$_6$ yet another specific feature is present, as documented in Fig. 4a. It is the negative curvature of the resistivity at high temperatures. This phenomenon found in many systems is usually attributed to the fact that the sample approaches the Mott limit [29, 30] at certain temperatures. In the case, the electron mean free path becomes comparable to the interatomic spacing. The effect is taken into account by introducing the empirical "parallel-resistor" with $\rho_{max}$ to the formula for resistivity $\rho$ [31]:

$$\frac{1}{\rho(T)} = \frac{1}{\rho_{BG}(T) + \rho_{BG}(0)} + \frac{1}{\rho_{max}}. \quad (5)$$

For fits of the resistivity data we have not used a standard BG formula but following the procedure used by Lortz we decomposed the spectral electron-phonon scattering function, $\alpha_{tr}^2 F(\omega)$, into a basis of Einstein modes each with a characteristic temperature $\theta_{E,k}$ ($k_B \theta_{E,k} = \hbar \omega_{E,k}$). Then, the discrete version of the generalized Bloch-Grüneisen formula is:

$$\rho_{BG}(T) = \frac{2\pi}{\varepsilon_0 \Omega_p^2} \sum_k \lambda_{tr,k} \theta_{E,k} \frac{x_k e^{x_k}}{(e^{x_k} - 1)^2}, \quad (6)$$

where $\Omega_p \equiv (ne^2/\varepsilon_0 m^*)^{1/2}$ is the unscreened plasma frequency, $\lambda_{tr,k} = (\alpha_{tr}^2 F)_k / \omega_k$ is partial contribution to the electron-phonon coupling function $\lambda$ (the dimensionless constant) and $\theta_{E,k}$ ($x_k = \theta_{E,k}/T$) is the Einstein temperatures of the phonon mode $k$.

Lortz et al. fitted their resistivity data by 3 modes with $\theta_{E,1} = 51$ K, $\theta_{E,2} = 90$ K, $\theta_{E,3} = 844$ K, and $\lambda_{tr,1} = 0.18$, $\lambda_{tr,2} = 0.72$, $\lambda_{tr,3} = 0.1$, respectively, i.e. with a dominant contribution of the Y vibration mode at 90 K (7.7 meV). In the first step we fitted our data at ambient pressure by the BG formula with the same set of modes and our residual resistivity $\rho(0) = 21$ μΩ.cm. The only fitting parameter was the resistivity of the parallel resistor for which we got a value $\rho_{max} = 140$ μΩ.cm. The resulting fit was of the same quality as the fit to the data of Lortz (Fig. 7 in Ref. [5]). In the following fits at all pressures we let the obtained value of $\rho_{max}$ fixed to minimize the number of fitting parameters. A value of the parallel resistor is related to the interatomic spacing or the lattice constant and since the latter parameter is changed by less than 1 % in the pressures up to 28 kbar (Fig. 3a) the value of $\rho_{max}$ should not change significantly, either.

To further minimize the number of fitting parameters, we tried to fit our data by the BG formula with just two modes of $\alpha^2 F(\omega)$, one located at low and one at high energy. This attempt was inspired by the point-contact spectra [10], where we could resolve only a single peak of the $\alpha^2 F(\omega)$ spectrum located at 7.5 meV ($\theta_E = 87$ K). Moreover, in the deconvolved spectrum of Lortz et al. near the main mode with $\theta_{E,2} = 90$ K and the weight $\lambda_{tr,2} = 0.72$, only a small contribution was found with $\theta_{E,1} = 51$ K and $\lambda_{tr,1} = 0.18$. Although a similar lowest-energy contribution has also been found by the ab initio calculations of the $\alpha^2 F(\omega)$ spectrum at ambient pressure [12], this mode is suppressed at higher pressure. Thus, we joined the modes $k = 1$ and $k = 2$ into a single mode denoted as $k = 1+2$. The resulting fit of the data at ambient pressure is shown in Fig. 4b. Besides the experimental points and the overall fit (solid line), contributions to the $d\rho(T)/dT$ dependence from the individual modes are also shown by the dashed lines. Indeed, the position of the maximum of the derivative $d\rho(T)/dT$ is fully determined by the low-energy mode(s). The quality of the resulting fit is not worse than of the one with 3 modes. This indicates limitations of such "thermal" spectroscopy, which can certainly give qualitative information, here for example on the importance of the low energy phonon modes, but cannot provide an exact spectral function with high energy resolution. The fitting parameters in the two-mode fit are $\theta_{E,1+2}$ = 87.4 K and $\theta_{E,3}$ = 844 K, and $\lambda_{tr,1+2}$ = 0.9 and $\lambda_{tr,3}$ = 0.1, respectively, for ambient pressure.

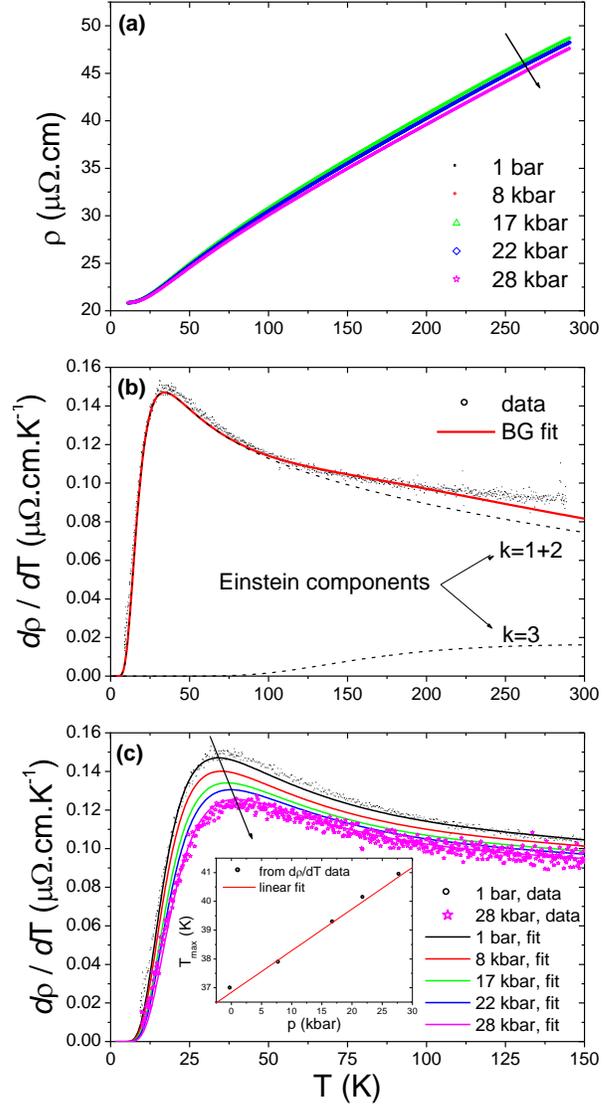

FIG. 4. Temperature dependences of the resistivity at 1 bar, 8, 17, 22, and 28 kbar for sample #2 (a) and its temperature derivative for 1 bar with the resulting Bloch-Grüneisen (BG) fit, which can be decomposed into two dominant Einstein terms (b). Temperature dependences of temperature derivatives of experimental data and corresponding BG fits for all pressures are described in (c). In the inset of (c) is shown pressure dependence of the temperature, at which maxima of $d\rho(T)/dT$ vs. $T$ are observed.

In the following we fit the $d\rho(T)/dT$ data taken at higher pressures. Since our main concern was the behavior of the low-temperature maximum, i.e. its shift to higher temperatures with increasing pressure, we let the values of $\theta_{E,3}$ = 844 K and $\lambda_{tr,3}$ = 0.1 be fixed and the only fitting parameters were $\theta_{E,1+2}$ and $\lambda_{tr,1+2}$. The generated $d\rho(T)/dT$ curves fit the data well, as demonstrated in Fig. 4c for the 1 bar and 28 kbar results. The main mode undergoes a smooth shift from $\theta_{E,1+2}$ = 87.4 K at ambient pressure to 102 K at 28 kbar, while the particular coupling constant contribution decreases from $\lambda_{tr,1+2}$ = 0.9 to 0.76. Those results can be compared with the calculations of Xu et al. [12]. They have calculated the Eliashberg function of the electron-phonon interaction $\alpha^2F(\omega)$ in the energy range up to 200 meV. For our purposes the most important result is that at ambient pressure the low energy part of the spectrum comprises the main peak at around 8 meV with a minor feature at around 7 meV. This spectrum is stiffened at 300 kbar. The minor feature is completely missing while the main peak is shifted to about 14 meV, i.e. to twice higher energy. In the linear response theory they calculated a increase of the logarithmically averaged frequency of the overall phonon modes shifting from 7 meV to 18.5 meV and then 24 meV for the pressures of 0, 200, and 400 kbar, respectively. The overall electron-phonon coupling $\lambda$ shifts from 1.44 down to 0.44 between 0 and 400 kbar and with the Coulomb pseudopotential $\mu^*$=0.1 it leads to a reduction of $T_c$ from 8.9 K to 1.9 K between 0 and 400 kbar. Within the rigid-muffin-tin approximation they have also calculated the square root of the weighted mean square of the phonon frequency from Y, $<\omega^2>_Y^{1/2}$ which can be compared with our low energy phonon mode shift since it originates just from the yttrium vibrations. The comparison is presented in Fig. 5. Even if the pressure ranges are very different for our data obtained from experiment and the calculations of Xu et al., one can see that the measured data fit quite well to the faster stiffening of the yttrium phonon mode calculated between 0 and 200 kbar. Later the calculated $<\omega^2>_Y^{1/2}$ yttrium phonons shift to higher frequencies more slowly. This tendency is in agreement with the changes of $T_c$, $H_{c3}$ and the sample volume upon increasing the pressure as presented in Sec. III.2. where initial faster change is gradually decelerated.

Within our simplified two-mode model of the electron-phonon interaction we also calculated the evolution of the transition temperatures with the applied pressure by the McMillan formula (2). Taking $\mu^*$= 0.1 and the $\theta_{E,k}$ and $\lambda_{tr,k}$ values we arrived to $T_c$ = 7.64 K and 7.12 K, for 1 bar and 28 kbar, respectively. The respective experimental values are 5.9 K and 4.8 K for these pressures. We can conclude that qualitatively our analysis based on the "thermal" spectroscopy yields a proper explanation for the suppression of superconductivity in YB$_6$ under pressure, which lies in the stiffening of the relevant yttrium phonon mode.

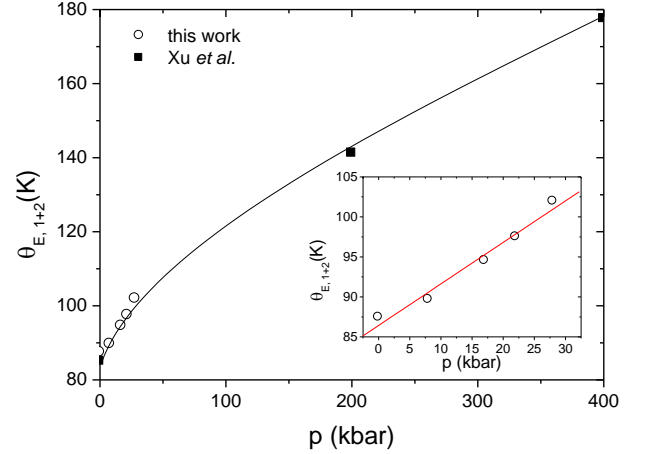

FIG. 5. Pressure dependence of the main phonon mode energy, together with theoretical data of Xu et al. [12] up to 400 kbar. The inset displays the detail of our measurements up to 28 kbar. Lines are guides for the eye.

## IV. CONCLUSIONS

Our results on the influence of hydrostatic pressure on magnetoresistivity up to 47 kbar show a decrease of the transition temperature $dT_c/dp$ = -0.044 K/kbar and the third critical field $dH_{c3}/dp$ = -4.64 mT/kbar. The measurements of magnetic susceptibility up to 112 kbar and lattice parameter up to 320 kbar show evidence that the $T_c$-reduction rate is slowed down at higher pressures. The *pressure versus volume* dependence can be described by the Rose-Vinet equation of state with the bulk modulus $B_0$ = 1659 kbar. The relative volume dependence of $d\ln T_c/d\ln V$ analyzed within the McMillan formula for $T_c$ indicates that the suppression of the superconducting transition temperature in YB$_6$ is mainly due to hardening of the Einstein-like phonon mode responsible for superconducting coupling. The precise zero-field resistivity measurements in the normal state between 10 and 300 K and up to 28 kbar used as a "thermal" spectroscopy corroborated that the low-energy Einstein-like phonon mode significant for superconductivity in YB$_6$ shifts to higher energy with pressure. Therefore, a reduction of the electron-phonon coupling constant leads to significant suppression of the superconducting transition temperature.

## ACKNOWLEDGEMENTS

We acknowledge L. Havela for careful reading of the manuscript and R. Hlubina for useful discussion. This work was supported by the project VEGA 2/0135/13, VEGA 2/0106/13, APVV 0036-11, APVV 0132-11, APVV-VVCE 0058, CFNT MVEP - the Center of Excellence of the Slovak Academy of Sciences, 7th FP EU-Microkelvin, nanoSC COST, the French National Research funding agency ANR project PRINCESS, EU ERDF-ITMS 26220120005, EU ERDF-ITMS26110230034 and EU ERDF-ITMS26110230097. The liquid nitrogen for the experiment was sponsored by the U.S. Steel Kosice, s.r.o.